\documentclass{PoS}

\usepackage{amsmath,amssymb}
\usepackage{graphicx}

\usepackage{setspace}
\usepackage{color}

\title{Multi-boson spectrum of the SU(2)-Higgs model}

\ShortTitle{SU(2)-Higgs Spectrum}

\author{\speaker{Mark Wurtz}\\
        Department of Physics and Astronomy, York University, Toronto, Ontario, M3J 1P3, Canada\\
        E-mail: \email{mwurtz@yorku.ca}}

\author{Randy Lewis\\
        Department of Physics and Astronomy, York University, Toronto, Ontario, M3J 1P3, Canada\\
        E-mail: \email{randy.lewis@yorku.ca}}

\abstract{Lattice simulations are used to compute the spectrum of energy levels for all angular momentum and parity quantum numbers in the SU(2)-Higgs model, with parameters chosen to match experimental data from the Higgs-$W$ boson sector of the standard model. Creation operators are constructed for all lattice irreducible representations, and a correlation matrix is formed from which the spectrum is extracted using a variational analysis. Many multi-boson states are observed and careful analysis reveals that all are consistent with weakly-interacting Higgs and $W$ bosons.}

\FullConference{31st International Symposium on Lattice Field Theory - LATTICE 2013\\
		July 29 - August 3, 2013\\
		Mainz, Germany}

\begin{document}

\section{Introduction}\label{sec:intro}

As underscored by the recent discovery of a 125 GeV Higgs-like particle [1,2], the standard model is an excellent description of nature up to presently-accessible energy scales. The Higgs-gauge sector of the standard model, with photons neglected, is described by the SU(2)-Higgs model, which is a useful effective field theory when a finite physical cutoff is in place. 

The SU(2)-Higgs model comprises both a confinement region and a Higgs region. The confinement region contains a rich spectrum of bound states reminiscent of QCD. The Higgs region is generally expected to contain only the Higgs and $W$ bosons, although there have been recent discussions of possible additional states in the Higgs sector \cite{Maas:2012tj,Maas:2012zf}.

The remainder of this article describes the lattice computations of Ref.~\cite{Wurtz:2013ova} where the Higgs-region spectrum was thoroughly studied with parameters tuned to match the standard model. More than a dozen energy levels were observed, all of which are consistent with the traditional expectation: weakly-interacting multi-particle states of Higgs and $W$ bosons.

\section{Operators}\label{sec:operators}

The SU(2)-Higgs Lagrangian contains gauge-dependent scalar $\phi(x)$ and gauge fields $U_\mu(x)$, but in the absence of gauge fixing, these do not have an obvious one-to-one connection with the physical particle states of the spectrum. Rather, the physical states couple to gauge-invariant operators that are composites of the fields. The Higgs boson, with $I(J^P)=0(0^+)$, where $I$ is the weak isospin, couples to the operator $\operatorname{Tr}\left(\phi^\dag(x)\phi(x)\right)$, which contains two scalar fields. The $W$ boson, $1(1^-)$, couples to the isovector gauge-invariant link $\operatorname{Tr}\left(-i\sigma^a\phi^\dag(x)U_\mu(x)\phi(x+\hat{\mu})\right)$.

Due to the discrete rotational symmetries of the lattice, angular momentum subduces to irreducible representations (irreps) $\Lambda$ of the octahedral group, as shown in Table~\ref{irrep_spin_table}. Therefore, the $0(0^+)$ Higgs and $1(1^-)$ $W$ correspond to $0(A_1^+)$ and $1(T_1^-)$, respectively.

\begin{table}[htb]
\caption{The number of copies of each irreducible representation $\Lambda$
         for each continuum spin $J$.}
\label{irrep_spin_table}
\centering{
\begin{tabular}{l|cccccccc}
$\Lambda$ & \multicolumn{8}{c}{$J$} \\
\cline{2-9}
 & 0 & 1 & 2 & 3 & 4 & 5 & 6 & $\dots$ \\
\hline 
$A_1$ & 1 & 0 & 0 & 0 & 1 & 0 & 1 & $\dots$ \\
$A_2$ & 0 & 0 & 0 & 1 & 0 & 0 & 1 & $\dots$ \\
$E$   & 0 & 0 & 1 & 0 & 1 & 1 & 1 & $\dots$ \\
$T_1$ & 0 & 1 & 0 & 1 & 1 & 2 & 1 & $\dots$ \\
$T_2$ & 0 & 0 & 1 & 1 & 1 & 1 & 2 & $\dots$
\end{tabular}
}
\end{table}

\begin{figure}[htb]
\centering
\includegraphics[scale=0.45]{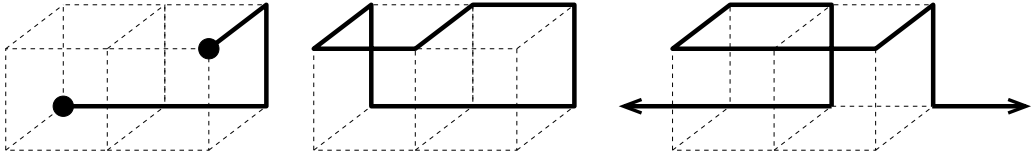}
\caption{Sketches of the gauge-invariant link, Wilson loop and Polyakov loop operators used to study the spectrum for all $I(\Lambda^P)$ channels.}
\label{figure_operators}
\end{figure}

To study the entire low-lying spectrum, zero-momentum operators with all possible $I(\Lambda^P)$ quantum numbers are constructed. The gauge-invariant link operator is composed of two scalar fields that are spatially separated and connected by gauge links, and is used to construct isoscalar ($I=0$) and isovector ($I=1$) operators. The Wilson loop and Polyakov loop operators are also employed to study all $\Lambda^P$ channels for $I=0$. Intricate shapes are chosen for the gauge-invariant link, Wilson and Polyakov loops, as shown in Fig.~\ref{figure_operators}, to allow access to all irreps and parity. Further details can be found in \cite{Wurtz:2013ova}.

Stout link and gaussian scalar field smearing techniques \cite{Morningstar:2003gk,Peardon:2009gh} were used to improve the operators and generate a large basis for a variational analysis. The number of stout link and scalar smearing iterations used for the variational basis are 0, 5, 10, 25, 50, 100, 150 and 200. It is worth noting that a total of 3840 operators were used in the analysis of the spectrum. Two isosinglet links, two isotriplet links, the Wilson loop and the Polyakov loop add to 10 operators. For each of the 10, we use all 24 spatial orientations, both parities, and 8 smearings. This gives (10)(24)(2)(8)=3840 operators just for zero momentum. We also studied several operators with non-zero momentum.

The energy spectrum is extracted by fitting to the exponential decay of correlation functions 
\begin{align}
C_{ij}(t) = \left< {\cal O}_i(t) {\cal O}_j(0) \right> &= \sum_n \left<0\right|{\cal O}_i\left|n\right> \left<n\right|{\cal O}_j\left|0\right> \exp\left(-E_n t \right) \\
&= \sum_n a_i^n a_j^n \exp\left(-E_n t \right)  \,\, ,
\end{align}
where ${\cal O}_i(t)$ are gauge-invariant operators with the vacuum expectation value subtracted. A variational method \cite{Wurtz:2013ova,Kronfeld:1989tb,Luscher:1990ck} is used to iteratively project out the lightest energies from the correlation matrix $C_{ij}(t)$:
\begin{align}
C_n(t) = z_n^i C_{ij}(t) z_n^j = A_n\exp\left(-E_nt\right) \,\, .
\end{align}

\section{Spectrum with a Physical Higgs Mass}\label{sec:spectrum}

\begin{figure}[tb]
\centering
\includegraphics[scale=0.45,clip=true]{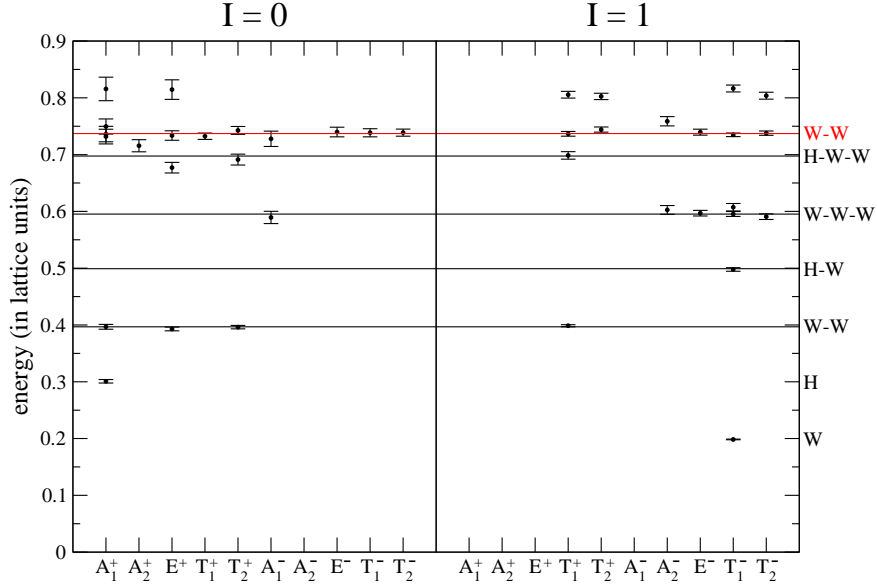}
\caption{Energy spectrum extracted from a variational analysis of correlation functions of gauge-invariant link, Wilson loop and Polyakov loop operators on a $20^3\times 40$ lattice with $\beta=8$, $\kappa=0.131$ and $\lambda=0.0033$. Data points are lattice results with statistical errors. Horizontal lines are expectations for non-interacting particles with (red) and without (black) internal momentum.}
\label{graph_mass_20x20x20x40}
\end{figure}

The low-lying mass spectrum for all $I(\Lambda^P)$, extracted using a variational analysis of correlation matrices of gauge-invariant operators, is shown in Fig.~\ref{graph_mass_20x20x20x40}. The lattice parameters are $\beta=8$, $\kappa=0.131$ and $\lambda=0.0033$ which result in a Higgs mass close to the physical value and a physical weak gauge coupling corresponding to $\frac{g^2}{4\pi}\approx\frac{\alpha}{\sin^2\theta_W}\approx 0.04$. The renormalized gauge coupling is close to the bare coupling value \cite{Langguth:1985dr}. The lattice size is $20^3\times 40$ with 20,000 configurations.

The Higgs and $W$ states are easily identified as the ground states in the $0(A_1^+)$ and $1(T_1^-)$ channels, respectively. The energy scale is set by defining the $W$ mass to be $80.4$ GeV, and the Higgs mass from our simulation is $122\pm1$ GeV. Of course a tiny adjustment of the $\lambda$ and $\kappa$ values would move the Higgs mass to exact agreement with experiment. Moving up in Fig.~\ref{graph_mass_20x20x20x40}, states with an energy of $2m_W$ appear in the $0(A_1^+)$, $0(E^+)$, $0(T_2^+)$ and $1(T_1^+)$ channels, completely consistent with the expected continuum quantum numbers for two stationary $W$ particles: $0(0^+)$, $0(2^+)$ and $1(1^+)$. This expectation follows from Bose statistics, which requires that the total wavefunction of the $W$-$W$ state, which contains spin and isospin components, be symmetric under interchange. No interaction energy is detectable within statistical errors at this weak gauge coupling. One step higher in Fig. 2, a state with an energy of $m_H+m_W$ is in the $1(T_1^-)$ channel, consistent with a stationary Higgs-$W$ pair which has the same quantum numbers as a single $W$. Next, states in the $0(A_1^-)$, $1(A_2^-)$, $1(E^-)$, $1(T_1^-)$ and $1(T_2^-)$ channels appear with an energy of $3m_W$, consistent with three stationary $W$ particles. Requiring that the three-$W$ state be symmetric under the interchange of any pair gives the follow continuum spins: $0(0^-)$, $1(1^-)$, $1(2^-)$ and $1(3^-)$, which completely agree with the channels listed above. Missing from the results in Fig.~\ref{graph_mass_20x20x20x40} is the two-Higgs state, expected to appear in $0(A_1^+)$ at an energy of $2m_H$. This state is observed in Sec.~\ref{Two-Particle Operators} using operators introduced in that section. Higher up are states with an energy of $m_H+2m_W$, found in the same channels as two stationary $W$'s, which is consistent with a Higgs and two $W$ particles all at rest.

The uppermost horizontal line in Fig.~\ref{graph_mass_20x20x20x40} (highlighted in red) represents the energy of two $W$ particles, each moving with the minimal momentum allowed on a lattice but total momentum equal to zero. The operators are defined to have zero total momentum but multi-particle states may appear which have internal momentum. In particular, two-particle states will have back-to-back momentum. The configuration of the internal momentum can not be specified by the lattice operators, because the momentum of each particle is not a conserved quantity. Momentum occurs on a lattice in integer multiples of $2\pi/L$ along the $x$, $y$ or $z$ axes, where $L$ is the spatial length of the lattice. Two moving particles with vanishing total momentum may also have orbital angular momentum. Again, not being a conserved quantity, the orbital angular momentum can not be specified by the lattice operators (much like the internal linear momentum). Only the total angular momentum, which corresponds to the lattice irrep $\Lambda$, is conserved. The orbital angular momentum can assume any value, and thus all $I(\Lambda^P)$ are expected for two moving $W$ bosons. This signal is observed in many channels, but not all. Section \ref{Two-Particle Operators} discusses why this signal did not appear in some channels. The data points at the top of Fig.~\ref{graph_mass_20x20x20x40} are difficult to interpret with the current simulation, but another simulation (discussed below) at a larger lattice volume can provide clarity.

\begin{figure}[tb]
\centering
\label{graph_mass_24x24x24x48}
\includegraphics[scale=0.45,clip=true]{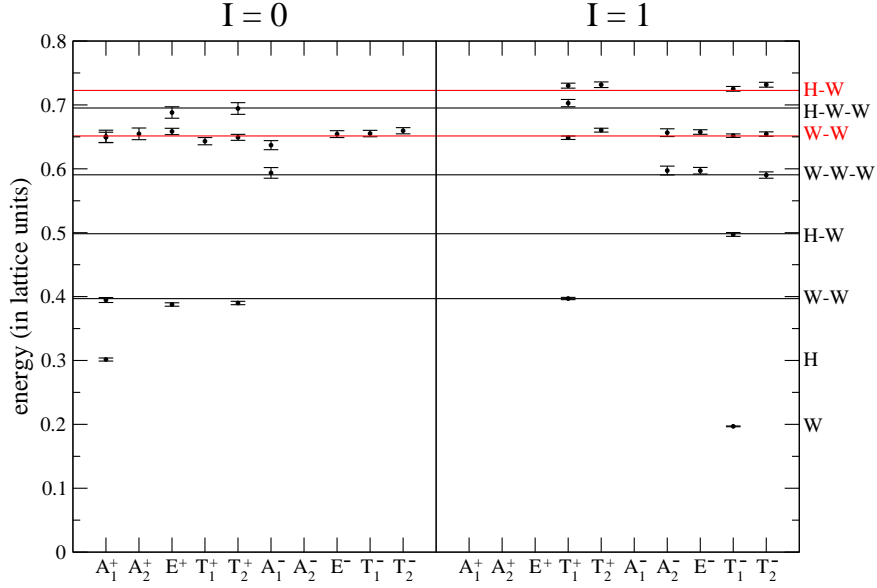}
\caption{The same as Fig.~\protect\ref{graph_mass_20x20x20x40} but on a (larger) $24^3\times 48$ lattice.}
\end{figure}

To verify the interpretation of states in Fig.~\ref{graph_mass_20x20x20x40} as two-particle states with the minimal non-zero back-to-to-back momentum, all lattice parameters will be held fixed ($\beta=8$, $\kappa=0.131$ and $\lambda=0.0033$) except the lattice size which is increased from $20^3\times 40$ to $24^3\times 48$. Increasing the lattice size decreases the minimal momentum $2\pi/L$. Spectrum results for the larger lattice are shown in Fig.~\ref{graph_mass_24x24x24x48}. The energy values remain unchanged except that those which have non-zero momentum (red horizontal lines in Fig.~\ref{graph_mass_24x24x24x48}) have decreased. Previously unidentified states from Fig.~\ref{graph_mass_20x20x20x40} are now identified as a Higgs-$W$ pair moving with the minimal momentum. While all $\Lambda^P$ with $I=1$ are expected to appear for a moving Higgs-$W$ pair (due to intrinsic orbital angular momentum), only a few channels observe a signal. Section~\ref{Two-Particle Operators} will discuss why there are missing irreps for multi-particle states with momentum.

\section{Two-Particle Operators}
\label{Two-Particle Operators}

Multi-particle operators are constructed by multiplying operators of definite momentum that predominantly couple to a single particle state. The "single" Higgs and $W$ operators, as a function of momentum $\vec{p}$, are given by
\begin{align}
H(\vec{p}) &=  \sum_{\vec{x}} \frac{1}{2} \operatorname{Tr}\left\{\phi^\dag(x)\phi(x)\right\} \, \exp\left\{i\vec{p}\cdot\vec{x}\right\} \,\, , \\
W^a_{\mu}(\vec{p}) &= \sum_{\vec{x}} \frac{1}{2} \operatorname{Tr}\left\{-i\sigma^a \phi^\dag(x) U_\mu(x) \phi(x+\hat{\mu})\right\} \, \exp\left\{i\vec{p}\cdot\left(\vec{x}+\tfrac{1}{2}\hat{\mu}\right)\right\}  \,\, .
\end {align}
"Two"-particle operators with back-to-back momentum are built as follows:
\begin{align}
&H(\vec{p})H(-\vec{p}) \,\, , \quad\quad  I=0 \,\, , \label{higgs-higgs} \\
&H(\vec{p})W^a_{\mu}(-\vec{p}) \,\, , \quad\,\,  \,\,  I=1 \,\, , \label{higgs-w} 
\end{align}
\begin{align}
&W^a_{\mu}(\vec{p})W^a_{\nu}(-\vec{p}) \,\, , \quad  I=0 \,\, , \label{ww0} \\
\epsilon^{abc}&W^b_{\mu}(\vec{p})W^c_{\nu}(-\vec{p}) \,\, , \quad  I=1 \,\, , \label{ww1}
\end{align}
where repeated indices $a$, $b$, $c$ are summed. Of course, these operators do not couple only to two particles with the specified internal momentum. They can, for example, couple to a single particle, two particles with a different momentum configuration, and more as long as the $I(\Lambda^P)$ quantum numbers are obeyed. However, this construction results in a much stronger contribution from two-particle states, which is needed to observe the two-Higgs state. All possible $I(\Lambda^P)$ are constructed from $H(\vec{p})$ and $W^a_{\mu}(\vec{p})$.

\begin{figure}[tb]
\centering
\includegraphics[scale=0.45,clip=true]{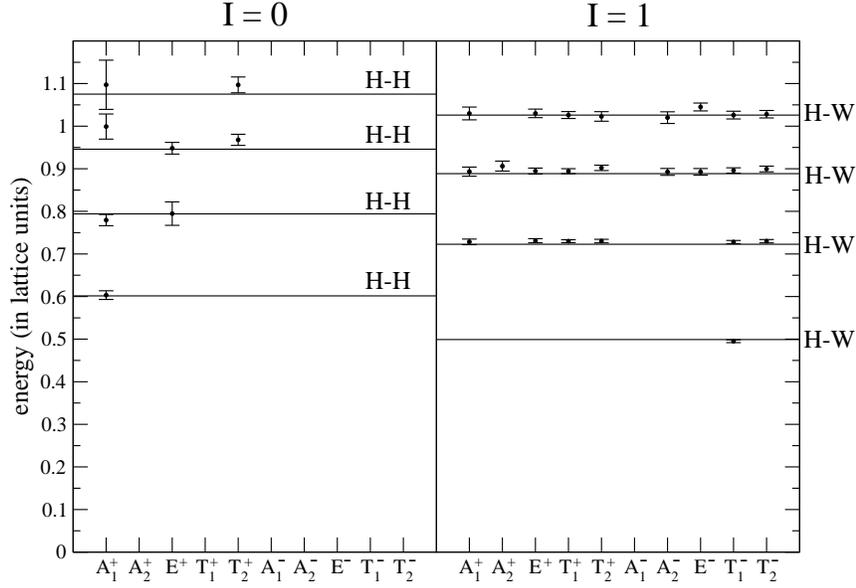}
\caption{Energy spectrum of Higgs-Higgs and Higgs-$W$ states with back-to-back momentum $\left|\vec{p}\right|=0$, $\left|\vec{p}\right|=2\pi/L$, $\left|\vec{p}\right|=\sqrt{2}(2\pi/L)$ and $\left|\vec{p}\right|=\sqrt{3}(2\pi/L)$. Lattice is $24^3\times 48$ with parameters $\beta=8$, $\kappa=0.131$ and $\lambda=0.0033$.}
\label{graph_mass_hh_hw}
\end{figure}

\begin{figure}[tb]
\centering
\includegraphics[scale=0.45,clip=true]{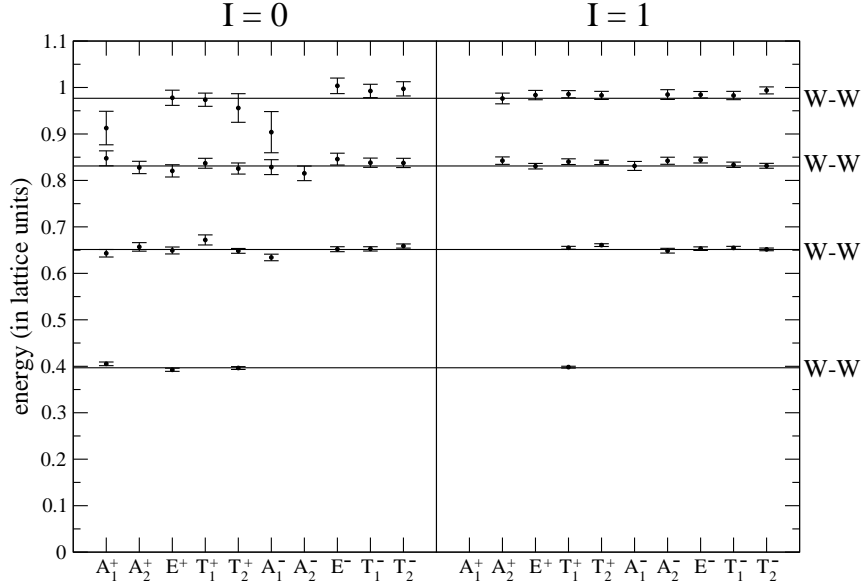}
\caption{Energy spectrum of $W$-$W$ states with back-to-back momentum $\left|\vec{p}\right|=0$, $\left|\vec{p}\right|=2\pi/L$, $\left|\vec{p}\right|=\sqrt{2}(2\pi/L)$ and $\left|\vec{p}\right|=\sqrt{3}(2\pi/L)$. Lattice is $24^3\times 48$ with parameters $\beta=8$, $\kappa=0.131$ and $\lambda=0.0033$.}
\label{graph_mass_ww}
\end{figure}

Figure~\ref{graph_mass_hh_hw} shows the Higgs-Higgs and Higgs-$W$ spectrum with back-to-back momenta $\left|\vec{p}\right|=0$, $\left|\vec{p}\right|=2\pi/L$, $\left|\vec{p}\right|=\sqrt{2}(2\pi/L)$ and $\left|\vec{p}\right|=\sqrt{3}(2\pi/L)$. The two-Higgs state is now easily found. The same calculation is performed for two $W$ particles in Fig.~\ref{graph_mass_ww}. There are a number of data points missing in Figs.~\ref{graph_mass_hh_hw} and \ref{graph_mass_ww} for two-particle states with non-zero momentum. The $W$-$W$ states with $\left|\vec{p}\right|=2\pi/L$ in Fig.~\ref{graph_mass_ww} match those found in Figs.~\ref{graph_mass_20x20x20x40} and \ref{graph_mass_24x24x24x48}. In fact, it is not possible to construct lattice operators of the form in Eqs.~\eqref{higgs-higgs}-\eqref{ww1} for the missing $I(\Lambda^P)$. Due to the lattice rotational symmetries, the direction of the internal momentum for multi-particle states restricts the allowed irreducible representations.

\section{Conclusions}\label{sec:conclusions}

The entire SU(2)-Higgs energy spectrum has been studied with all parameters tuned to match the standard model. The multi-boson spectrum was observed and is consistent with collections of weakly interacting Higgs and $W$ bosons.

\section*{Acknowledgments}

The authors thank Colin Morningstar for helpful discussions about the smearing of
lattice operators.
This work was supported in part by the Natural Sciences and
Engineering Research Council (NSERC) of Canada, and by computing resources of
WestGrid\cite{westgrid} and SHARCNET\cite{sharcnet}.

\end{document}